# Gravitomagnetic Field of a Rotating Superconductor and of a Rotating Superfluid

M. Tajmar[*]

*ARC Seibersdorf research GmbH, A-2444 Seibersdorf, Austria*

C. J. de Matos[†]

*ESA-ESTEC, Directorate of Scientific Programmes, PO Box 299, NL-2200 AG Noordwijk, The Netherlands*

**Abstract**

The quantization of the extended canonical momentum in quantum materials including the effects of gravitational drag is applied successively to the case of a multiply connected rotating superconductor and superfluid. Experiments carried out on rotating superconductors, based on the quantization of the magnetic flux in rotating superconductors, lead to a disagreement with the theoretical predictions derived from the quantization of a canonical momentum without any gravitomagnetic term. To what extent can these discrepancies be attributed to the additional gravitomagnetic term of the extended canonical momentum? This is an open and important question. For the case of multiply connected rotating neutral superfluids, gravitational drag effects derived from rotating superconductor data appear to be hidden in the noise of present experiments according to a first rough analysis.

---

[*] Research Scientist, Space Propulsion, Phone: +43-50550-3142, Fax: +43-50550-3366, E-mail: martin.tajmar@arcs.ac.at

[†] Scientific Advisor, Phone: +31-71-565-3460, Fax: +31-71-565-4101, E-mail: clovis.de.matos@esa.int

**Introduction**

Applying an angular velocity $\vec{\omega}$ to any substance aligns its elementary gyrostats and thus causes a magnetic field known as the *Barnett effect* [1]. In this case, the angular velocity $\vec{\omega}$ is proportional to a magnetic field $\vec{B}_{equal}$ which would cause the same alignment:

$$\vec{B}_{equal} = -\frac{1}{g}\frac{2m}{e} \cdot \vec{\omega}. \qquad (1)$$

where $g$ is the Landé factor, $m$ and $e$ are the mass and the charge of the electron respectively. As recently shown by the authors [2], a gravitational analogue exists using the weak-field approximation of general relativity, in which the field equations can be written in a form similar to Maxwell's equations with a gravitoelectric field $\vec{g}$ [m.s$^{-2}$] analogue to the electric field $\vec{E}$ [V.m$^{-1}$] and a gravitomagnetic field $\vec{B}_g$ [rad.s$^{-1}$] analogue to the magnetic field $\vec{B}$ [T] [3, 4]. In this so called *gravitomagnetic Barnett effect*, the applied angular velocity is also equivalent to a gravitomagnetic field $\vec{B}_{g,equal}$ which results in a gravitomagnetization of the substance:

$$\vec{B}_{g,equal} = -\frac{2}{g} \cdot \vec{\omega}. \qquad (2)$$

However, since the gravitomagnetic permeability $\mu_g = 4\pi G/c^2 = 9.31 \times 10^{-27}$ m.kg$^{-1}$ is so small, the resulting gravitomagnetic field outside the substance is too weak to be detected. This low permeability leads to extremely small order of magnitude effects for laboratory type experiments on general relativity [5].

We propose in this paper, that a different coupling between electromagnetism and gravito-electromagnetism might exist in quantum materials such as superconductors or superfluids.

**Rotating Superconductors**

Let us first consider a superconductor in the form of a cylindrical shell. According to London [6], the local mean value of the canonical momentum vector of the superelectrons $\vec{p}_s$, integrated around a closed path, is quantized:

$$\oint \vec{p}_s \cdot d\vec{l} = \oint (m\vec{v}_s + e\vec{A}) \cdot d\vec{l} = \frac{nh}{2}, \qquad (3)$$

where $m$ and $\vec{v}_s$ are respectively the mass and the velocity of the Cooper-pairs, $\vec{A}$ is the magnetic vector potential, $n$ is an integer and $h$ is the Planck's constant. London argued [6], that this quantization condition must vanish for a closed loop inside the superconductor giving $n=0$. Taking the curl of Equ. (3) and London's argument, we see that a rotating superconductor with an angular velocity $\vec{\omega}$ will produce a magnetic field $\vec{B}$ given by

$$\vec{B} = -\frac{2m}{e} \cdot \vec{\omega}. \qquad (4)$$

This is also called the London moment, arising from Coriolis forces on a charged particle in a rotating reference frame. More generally, this expression can be derived within the framework of the Ginzburg-Landau theory, integrating the current density equation around a closed path including the effect of the rotating reference frame [7]. We then obtain

$$\frac{m}{e^2 n_s} \oint_\Gamma \vec{j} \cdot d\vec{l} = \frac{nh}{2e} - \int_{S_\Gamma} \vec{B} \cdot d\vec{S} - \frac{2m}{e} \cdot \vec{\omega} \cdot \vec{S}_\Gamma, \qquad (5)$$

where $n_s$ is the Cooper-electron number density and $S_\Gamma$ is the area bounded by $\Gamma$.

If the superconducting ring is thick with respect to the London penetration depth, we can find a contour $\Gamma$ within the superconductor many penetration depths away where the current density $\vec{j}$, and hence $n$ according to London, is zero. Equ. (5) then reduces to the classical London moment Equ. (4). The magnetic field produced by such superconductors has been verified by a number of experiments [8-10], also for high-$T_c$ and heavy-fermion superconductors, matching Equ. (2) within a few percent.

If the superconducting ring is thin compared to the London penetration depth, the current density $\vec{j}$ will be constant and Equ. (4) does not apply any more. There will then be an angular velocity $\omega_n$ for each $n$ so that $\vec{j}$ and $\vec{B}$ will be zero (assuming that the magnetic field is solely caused by the Cooper-electron current). We can then write the zero flux condition as

$$\frac{h}{2\pi m} = 2 S_\Gamma \Delta \nu, \qquad (6)$$

where $\Delta \nu$ is the frequency between each $n$. This equation can be used to determine the Cooper-pair mass $m$. In the most accurate experiment up to now, Tate et al [11, 12] used a Niobium superconducting ring with a SQUID device, measuring $(m/2m_e)=1.000084(21)$ where $m_e$ is the mass of a free electron, and the errors from systematic effects were estimated at 21 ppm. This result is accurate enough to compare with theoretical predictions of 0.999992, taking into account a large kinetic energy term for electrons near the Fermi surface and a contribution to the magnetic vector potential within the superconductor from the motion of the internal electrostatic potential in the laboratory frame. This disagreement with theory is discussed in the literature [13], without any apparent solution.

We now propose to extend Equ. (3) and Equ. (5) to include also gravitational effects [14, 15] using the gravitomagnetic vector potential $\vec{A}_g$ and gravitomagnetic Field $\vec{B}_g$,

$$\oint \vec{p}_s \cdot d\vec{l} = \oint \left( m\vec{v}_s + e\vec{A} + m\vec{A}_g \right) \cdot d\vec{l} = \frac{nh}{2}, \tag{7}$$

$$\frac{m}{e^2 n_s} \oint_\Gamma \vec{j} \cdot d\vec{l} = \frac{nh}{2e} - \int_{S_\Gamma} \vec{B} \cdot d\vec{S} - \frac{m}{e} \int_{S_\Gamma} \vec{B}_g \cdot d\vec{S} - \frac{2m}{e} \cdot \vec{\omega} \cdot \vec{S}_\Gamma. \tag{8}$$

In the case of a thick superconducting ring ($\vec{j}$ and $n$ equal to zero inside), we obtain a simple modification to London's moment,

$$\vec{B} = -\frac{2m}{e} \cdot \vec{\omega} - \frac{m}{e} \cdot \vec{B}_g. \tag{9}$$

To know how much the gravitomagnetic field influences the generation of a magnetic field and vice versa, we need more information about the ratio between $\vec{B}$ and $\vec{B}_g$. The classical coupling between the two fields is given by the ratio of the gravitomagnetic and magnetic permeabilities and the electron's mass-to-charge ratio [3] given by $\vec{B}_g = -\frac{4\pi G}{c^2 \mu_0} \frac{m}{e} \vec{B} = -7.41x10^{-21} \cdot \frac{m}{e} \cdot \vec{B}$. Equ. (9) would then lead to

$$\vec{B} = -\frac{2m}{e \cdot \left(1 - 7.41x10^{-21}\right)} \cdot \vec{\omega} \cong -\frac{2m}{e} \cdot \vec{\omega}, \tag{10}$$

$$\vec{B}_g = -\frac{2\vec{\omega}}{\left(1 - 1.35x10^{20}\right)} = \vec{\omega} \cdot 1.48x10^{-20}. \tag{11}$$

In the classical coupling, the magnetic field would basically stay unaffected and the resulting gravitomagnetic field would be extremely small (for comparison, the gravitomagnetic field of the Earth on the equatorial plane is about $10^{-14}$ rad.s$^{-1}$). However, since the permeabilities do not enter in the quantization equation, the classical coupling may not be correct. In the case of Hildebrandt's experiment [8], from Equ. (9) a value of $B_g \approx 7,000$ rad.s$^{-1}$ would be needed to match the experimental observations due to the margin of a few percent in matching London's classical moment equation. Of course 7,000 rad.s$^{-1}$ is unrealistically high (the weak field approximation used in our approach would certainly break before), however, this upper boundary of such a field possibly involved needs to be tested (and certainly reduced) by subsequent experiments. Due to the high order of magnitude, this seems to be a worthwhile investigation.

By analysing thin superconducting experiments following Tate et al, we may find a different zero flux condition as the one in Equ. (6), because the gravitomagnetic flux not necessarily has to vanish at $\vec{j} = 0$ as the magnetic flux (which is assumed to originate from the Cooper-pair electron current only), it can also include contributions from the superconductor's neutral lattice structure. We then arrive at

$$\frac{h}{2\pi m} = 2S\left(\Delta v + \frac{\Delta B_{g,lattice}}{2\pi}\right). \tag{12}$$

For Tate's experiment, we see from Equ. (12) that a $\Delta B_{g\ lattice}=1.065\times10^{-8}$ rad.s$^{-1}$ would be required to match the experimental data. This is certainly above any classical coupling phenomena and needs to be investigated. Of course it can turn out that a refinement of the corrections to the Cooper-pair mass will give a better match between experiment and theory, however, at the present stage the investigation of the gravitomagnetic characteristics of a rotating superconductor (e.g. using spinning gyroscopes) seem to be an attractive experiment.

**Rotating Superfluids**

Due to the absence of charge, the quantization condition for superfluids including gravitomagnetic effects can be written as

$$\oint \vec{p}_s \cdot d\vec{l} = \oint \left( m\vec{v}_s + m\vec{A}_g \right) \cdot d\vec{l} = nh, \qquad (13)$$

where $m$ is the bare atomic mass in $^4$He or twice the atomic mass of $^3$He (due to Cooper pairing of atoms). Unfortunately, London's argument of $n=0$ in a closed loop does not hold in this case since the gravitomagnetic Meissner effect does not exist [16]. For a typical superfluid we will therefore find no contour in which the superfluid current will be zero (also the closed loop integration is not so simple due to the phenomena associated with vortices). In a gravitational context, a superfluid shall therefore behave similar to a very thin superconductor as in Tate's experiment with different $n$'s.

The effect could be studied in the context of phase-slip gyroscopes [17], where a Josephson weak link inside a vessel containing the superfluid is rotated. This causes a phase difference across this weak link and hence the superfluid will exist in a non-zero velocity state.

In a very recent experiment, Simmonds et al [18] report results on a double-path quantum interferometer using $^3$He, showing an interference pattern due to the Earth's rotation matching predictions within a systematic uncertainty of 10%. The Josephson weak link current in their experiment is characterized by

$$I_c^* = 2I_c \cdot \left|\cos\left(\pi \frac{2\omega S}{\kappa}\right)\right|, \qquad (14)$$

where $\kappa=h/(2m)$ with $m$ as the atomic mass of $^3$He. By following their derivation and including gravitomagnetic effects, we arrive at the modified equation

$$I_c^* = 2I_c \cdot \left|\cos\left[\pi \frac{2\omega S}{\kappa}\left(1+\frac{B_g}{2\omega}\right)\right]\right|. \qquad (15)$$

For a very rough estimate, taking the gravitomagnetic field derived from Tate ($B_g \approx 10^{-8}$ rad.s$^{-1}$) and the angular velocity of the Earth ($\omega_E = 7.2 \times 10^{-5}$ rad.s$^{-1}$), the phase would change by a factor of 1.00007 with respect to the measurements made by Simmonds et al. This is orders of magnitude below the accuracy of a few percent of Simmonds experiment, and hence, might not have been detected yet.

**Conclusion**

Present uncertainties in experiments with rotating superconductors and superfluids leave a very high upper boundary of possibly involved gravitomagnetic fields. This shall stimulate the investigation of the gravitomagnetic properties of such rotating superconductors and superfluids, for example by measuring the torque on a spinning gyroscope produced by the gravitomagnetic field possibly generated by rotating superconductors and superfluids. According to the knowledge of the authors, this experiment has not been done. According to our analysis, it could be a worthwhile task however.